\documentclass[aps,prd,rapid,nofootinbib]{revtex4}
\usepackage{graphicx}
\usepackage{epsfig}
\usepackage{slashed}

\newcommand{\bel}[1]{\begin{equation}\label{#1}}
\newcommand{\bal}[1]{\begin{eqnarray}\label{#1}}

\newcommand{\be}{\begin{equation}}
\newcommand{\ee}{\end{equation}}
\newcommand{\ba}{\begin{eqnarray}}
\newcommand{\ea}{\end{eqnarray}}

\newcommand{\bes}{\begin{equation*}}
\newcommand{\ees}{\end{equation*}}

\usepackage{latexsym}
\usepackage{amsmath}
\usepackage{amssymb}
\usepackage{eufrak}
\usepackage{euscript}
\usepackage{epsfig,graphics,graphicx}
\usepackage{bm}
\usepackage{color}
\usepackage{feynmf}
 \usepackage{slashed}

\def\be{\begin{eqnarray}}
\def\ee{\end{eqnarray}}

\def\Ep{E_{{\bf p}}}

\def\Eq{E_{{\bf q}}}

\begin{document}
\title{Effect of quark masses on the QCD presssure in a  strong  magnetic background}

\author{Jean-Paul Blaizot$^1$, Eduardo S. Fraga$^2$ and Let\'\i cia F. Palhares$^{1,2,3}$}

\affiliation{$^1$Institut de Physique Th\'eorique, CEA-Saclay, 91191 Gif-sur-Yvette, France \\
$^2$Instituto de F\'\i sica, Universidade Federal do Rio de Janeiro, 
Caixa Postal 68528, Rio de Janeiro, RJ 21941-972, Brazil \\  
$^3$ Institut f\"ur Theoretische Physik, Universit\"at Heidelberg, Philosophenweg 16, 69120 Heidelberg, Germany}

\begin{abstract}
We compute the 
two-loop contribution to the  QCD pressure in a strong magnetic background, for arbitrary quark masses. We show that,  for very large fields, the chiral limit is trivial.
\end{abstract}

\maketitle


Large magnetic fields 
can be created not only in the core of magnetars \cite{magnetars} but also in current experiments 
at BNL/RHIC and CERN/LHC involving non-central heavy ion collisions. The fields created 
in these collisions are possibly the largest magnetic fields  produced  since the primordial electroweak transition, reaching 
values $B \sim 10^{19}~$Gauss ($eB \sim6\,m_{\pi}^2$) for peripheral collisions at 
RHIC \cite{magnetic-HIC}, and even much higher at the LHC thanks to the fluctuations in the distribution 
of protons inside the nuclei \cite{Bzdak:2011yy}. 
Such intense magnetic fields may dramatically affect the phases of strongly interacting matter, as is the case in more ordinary circumstances \cite{landau-book}. The mapping of the  QCD
 phase diagram in the $T-eB$ plane  is still in its infancy (see e.g. \cite{Fraga:2012rr} and references 
therein). There are clear indications that sufficiently large magnetic fields do modify the nature and behavior 
of the chiral and the deconfinement phase transitions 
\cite{Agasian:2008tb,Fraga:2008qn,Avancini:2012ee,Boomsma:2009yk,Fukushima:2010fe,Johnson:2008vna,Preis:2010cq,Kashiwa:2011js,Chatterjee:2011ry,Andersen:2011ip,Andersen:2012dz,Skokov:2011ib,Fraga:2012fs,Fukushima:2012xw}. New phases are also predicted \cite{Ferrer:2005vd,Fukushima:2007fc,Noronha:2007wg,Son:2007ny}, and it has even been suggested that the vacuum 
may  turn into a superconducting medium via $\rho$-meson 
condensation \cite{Chernodub:2010qx}. 

While most of the analyses so far have relied on effective models,  or calculations in large $N_c$ limit of QCD \cite{Fraga:2012ev}, the first results from lattice QCD have been obtained recently \cite{lattice-maxim,lattice-delia,Bali:2011qj,Bali:2012zg}. This opens a new  channel for comparison between analytical or semi-analytical techniques and numerical non-perturbative 
approaches. In this perspective, we note that the recently established discrepancy between different lattice QCD results (for large \cite{lattice-delia} and physical values \cite{Bali:2011qj,Bali:2012zg} of quark masses) is most likely  related to quark mass effects. It is the purpose of this paper to analyze the possible competition between mass and magnetic-field corrections to the QCD pressure.
More specifically, we compute the two-loop correction to the  QCD pressure 
in a magnetic background field and for arbitrary quark masses. We indeed find a significant competition between the effects of quark masses and those of the magnetic background.
In particular,  for extremely intense magnetic fields, we show that the two-loop contribution  to the pressure is trivial in the chiral limit.

We shall assume in our calculation a constant 
and uniform Abelian magnetic background, whose strength is  large enough to produce interesting effects, i.e.  $eB\gtrsim m_{\pi}^2$. We also consider the temperature to be large 
enough that perturbation theory can be applied to the calculation of the pressure. 
Nevertheless, the effect of the strong background magnetic field  must be treated non-perturbatively: this is achieved by using the 
propagator  that was obtained long ago by Schwinger \cite{Schwinger:1951nm}, and that  can 
be cast in a  convenient form using  Landau levels, as shown in 
Ref. \cite{Chodos:1990vv} (see also Refs. \cite{Gusynin:1995nb,Fukushima:2011nu,leticia-tese}). 
Since we restrict our analysis to the case of very intense magnetic fields, the summation over the Landau levels is rapidly convergent, and the leading correction to the pressure is obtained from  the Lowest Landau Level (we shall refer to such calculations  as  the lowest Landau level  (LLL) approximation). The corresponding  propagator for a fermion of a given flavor $f$ and  (absolute) electric charge 
$q_{f}$,  in the presence of the classical field $A_{\rm cl}=(0,\vec{A})$ (with $\nabla\times \vec{A}=\vec{B}=B \hat z$) reads:
\begin{eqnarray}
S_0^{\rm LLL}(x,y)
&=&
{\rm exp}\left\{\frac{iq}{2}[x^{\mu}-y^{\mu}]A_{\mu}^{\rm ext}(x+y)\right\}
\int\frac{d^dP}{(2\pi)^d}
{\rm e}^{-iP\cdot(x-y)}
i {\rm exp}\left(-\frac{{\bf p}_T^2}{|qB|}\right)
~\frac{1+i\gamma^1\gamma^2}{{\bf p}_L\cdot\gamma_L-m_f}
\,,
\nonumber\\
\end{eqnarray}
where we have used a compact notation for the transverse  (${\bf p}_T=(p_1,p_2)$, ${\bf \gamma}_T=(\gamma^1,\gamma^2)$) and longitudinal (${\bf p}_L=(p_0,p_3)$,  ${\bf \gamma}_L=(\gamma^0,\gamma^3)$) quantities.
This is equivalent to the result used in Ref. \cite{Fukushima:2011nu}, obtained by constructing the projectors on the different Landau levels from the exact solution of the Dirac equation \cite{Fukushima-projectors}. A peculiarity of the LLL approximation should be noted: the propagator is  a $4\times 4$ matrix, but it describes only two physical propagating modes. Two eigenvalues of $S_{0}^{\rm LLL}$ indeed go to zero in the vicinity of the lowest Landau level pole ($p_0^2=m_f^2+p_3^2$). This complicates  in particular the computation of the free pressure in the LLL approximation \cite{Fraga:2012fs,leticia-tese}. 

The  thermodynamic 
potential of QCD, up to two loops, is obtained form the standard diagrammatic expansion:
\vspace{0.2cm}
\begin{fmffile}{fmftese}
\begin{eqnarray}
\Omega_{QCD}&\equiv& -~\frac{1}{\beta V}~\ln Z_{QCD}
\nonumber \\ \nonumber \\
&=& -~\frac{1}{\beta V}~~
\parbox{10mm}{
\begin{fmfgraph*}(35,35)\fmfkeep{bolhagluon}
\fmfpen{0.8thick}
\fmfleft{i} \fmfright{o}
\fmf{gluon,left,tension=.08}{i,o}
\fmf{gluon,left,tension=.08}{o,i}
\end{fmfgraph*}}
\quad
+~\frac{1}{\beta V}~~
\parbox{10mm}{
\begin{fmfgraph*}(35,35)\fmfkeep{bolhaghost}
\fmfpen{0.8thick}
\fmfleft{i} \fmfright{o}
\fmf{dashes,left,tension=.08}{i,o}
\fmf{dashes,left,tension=.08}{o,i}
\end{fmfgraph*}}
\quad
+\frac{1}{\beta V}~\sum_{f}~
\parbox{10mm}{
\begin{fmfgraph*}(35,35)\fmfkeep{bolhaquark}
\fmfpen{0.8thick}
\fmfleft{i} \fmfright{o}
\fmf{fermion,left,tension=.08,label=$\noexpand\psi_f$}{i,o}
\fmf{fermion,left,tension=.08}{o,i}
\end{fmfgraph*}}
\quad
+\nonumber \\ \nonumber \\ \nonumber\\
&&
+ ~\frac{1}{2}~\frac{1}{\beta V}~\sum_{f}~~
\parbox{10mm}{
\begin{fmfgraph*}(35,35)\fmfkeep{exchange}
\fmfpen{0.8thick}
\fmfleft{i} \fmfright{o}
\fmf{fermion,left,tension=.08,label=$\noexpand\psi_f$}{i,o,i}
\fmf{gluon}{i,o}
\fmfdot{i,o}
\end{fmfgraph*}}
~~~
+ ~\frac{1}{2}~\frac{1}{\beta V}~~
\parbox{10mm}{
\begin{fmfgraph*}(35,35)\fmfkeep{exchange-ghost}
\fmfpen{0.8thick}
\fmfleft{i} \fmfright{o}
\fmf{dashes,left,tension=.08}{i,o,i}
\fmf{gluon}{i,o}
\fmfdot{i,o}
\end{fmfgraph*}}
~~~
- ~\frac{1}{2}~\frac{1}{\beta V}\frac{1}{6}~~
\parbox{10mm}{
\begin{fmfgraph*}(35,35)\fmfkeep{exchange-3g}
\fmfpen{0.8thick}
\fmfleft{i} \fmfright{o}
\fmf{gluon,left,tension=.08}{i,o,i}
\fmf{gluon}{i,o}
\fmfdot{i,o}
\end{fmfgraph*}}
~~~
- 
\nonumber \\ \nonumber\\ \nonumber \\
&& -\frac{1}{2}~\frac{1}{\beta V}\frac{1}{8}~~~~~
\parbox{10mm}{
\begin{fmfgraph*}(35,35)\fmfkeep{db-4g}
\fmfpen{0.8thick}
\fmfbottom{i} \fmftop{o}
\fmf{phantom}{i,v}
\fmf{gluon,tension=.5}{v,v}
\fmf{phantom}{v,o}
\fmf{gluon,left=90,tension=.5}{v,v}
\fmfdot{v}
\end{fmfgraph*}}
~~~ +
\nonumber \\ \nonumber \\
&& +~ [diagrams ~with ~counterterms]~
+~O(3~loops) , \label{OmegaY}
\end{eqnarray}
\vspace{-0.4cm}

\noindent where full lines are fermions, dressed by the magnetic field, curly lines are gluons and dashed lines represent ghosts (whose role is essentially to  cancel the contribution of spurious degrees of freedom in the gluonic pressure). The calculation is carried out in Feynman gauge. 

The gluonic part is equivalent to the usual hot perturbative QCD result and is, therefore, 
well-known \cite{kapusta-gale}:
\be
\Omega_{QCD}^G
&=&
-2(N_c^2-1)\frac{\pi^2T^4}{90}
+(N_c^2-1)N_c~g^2T^4~\frac{1}{144}
\, .
\label{OmegaGQCD}
\ee

The one-loop contribution to the fermionic pressure has been considered in different contexts (usually, in effective field theories \cite{Fraga:2008qn,Boomsma:2009yk,Avancini:2012ee,Andersen:2011ip,Fraga:2012fs,Ebert:2003yk}) and computed from the direct knowledge of the Landau levels 
$E^2(n,p_3)=p_3^2+m_f^2+2q_fBn$ and their degeneracies $q_fB/(2\pi)$ for $n=0$ and 
$q_fB/\pi$ for $n=1,2,\cdots$. 
The final exact result reads  
(see Ref. \cite{Fraga:2012fs} for discussions on the subtraction procedure)
\be
\frac{P_{\rm free}^{F}}{N_c}&=&
\sum_f \frac{(q_fB)^2}{2\pi^2}\Big[
\zeta'\left(-1,x_f\right)-\zeta'\left(-1,0\right)
+\frac{1}{2}(x_f-x_f^2)\ln x_f+\frac{x_f^2}{4}
\Big]
\nonumber\\&
+&T \sum_{n,f}\frac{q_fB}{\pi}(1-\delta_{n0}/2)
\int \frac{dp_3}{2\pi} \bigg\{ 
\ln\left( 1+e^{-\beta[E(n,p_3)-\mu_f]} \right)
+\ln\left( 1+e^{-\beta[E(n,p_3)+\mu_f]} \right)
\bigg\}
\,,
\nonumber\\
\ee
where $\mu_f$ is the quark chemical potential (associated to baryon number conservation). In the limit of large magnetic field (i.e. $x_f=m_f^2/(2q_fB)\to 0$), reduces to the LLL contribution
\be
\frac{P_{\rm free}^{F}}{N_c}&\stackrel{{\rm large}~ B}{=}&
\sum_f \frac{(q_fB)^2}{2\pi^2}\Big[
x_f\ln \sqrt{x_f}
\Big]
+T \sum_{f}\frac{q_fB}{2\pi}
\int \frac{dp_3}{2\pi} \bigg\{ 
\ln\left( 1+e^{-\beta[E(0,p_3)-\mu_f]} \right)
+\ln\left( 1+e^{-\beta[E(0,p_3)+\mu_f]} \right)
\bigg\}
\,.
\nonumber\\
\ee

The exchange diagram corresponds to the first nontrivial contribution. In terms of the propagators in coordinate space, this diagram is given by

\begin{eqnarray}
\parbox{10mm}{
\fmfreuse{exchange}
}
\quad &=& 
\beta V~g^2~N_c(\lambda_a\lambda_a)~
\int\frac{d^dxd^dy}{\beta V}\int\frac{d^dK}{(2\pi)^d}
\frac{{\rm e}^{-iK\cdot(y-x)}}{K^2}
{\rm Tr}
\big[
\gamma_{\mu}S_0(x,y)\gamma^{\mu}S_0(y,x)
\big]
\, , \nonumber\\ \label{excRF}
\end{eqnarray}
\noindent where $\lambda_a$ are Gell-Mann matrices, with $\lambda_a \lambda_a=(N_c^2-1)/2$,
the trace $\textrm{Tr}$ acts over Dirac indices  and the 4-momentum is given in terms of the
Matsubara frequencies ($\omega_l^B=2l\pi T$) 
and of the 3-momentum ${\bf k}$  as: $K=\left( k^0=i\omega_l^{B}\, ,\, {\bf k} \right)$.

Notice that if we assume translational invariance and the free Dirac propagator for the fermions (with $P=\left( p^0=i\omega_{n}^{F}+\mu_f \, ,\,  {\bf p}\right)$ and $\omega_n^F=(2n+1)\pi T$), 
this expression reduces to the usual one \cite{kapusta-gale}. In the presence of a uniform and constant magnetic background (${\bf B}=B\hat z$), however, the fermion propagator becomes dependent on $x$ and $y$ in a nontrivial way due to the Schwinger phase, as discussed previously.

However, a detailed analysis of this diagram shows that it can be cast in the following neat form \cite{leticia-tese}:
\begin{eqnarray}
\parbox{10mm}{
\fmfreuse{exchange}
}^{\quad\rm LLL}
 = 
\left(
\frac{q_fB}{2\pi}\right)
\int\frac{dk_1dk_2}{(2\pi)^2}~{\rm e}^{
-\frac{k_1^2+k_2^2}{2q_fB}}~
\mathcal{G}\left(k_1^2+k_2^2,m_f^2\right)
 &=& 
\left(
\frac{q_fB}{2\pi}\right)
\int\frac{d {\bf k}_{\perp}}{(2\pi)^2}~{\rm e}^{
-\frac{{\bf k}_{\perp}^2}{2q_fB}}\quad
\parbox{10mm}{
\fmfreuse{exchange}
}^{\quad  \bar d=2}_{\; \;\;m_k^2={\bf k}_{\perp}^2}
,
 \label{excRF-LLL-4}
\end{eqnarray}

\vspace{.5cm}

\noindent 
where ($\nu_c\equiv N_c(N_c^2-1)/2$)
\begin{eqnarray}
\mathcal{G}\left(m_k^2,m_f^2\right)
&=&
\beta Vg^2\nu_c
\int\frac{dk_zdp_zdq_z}{(2\pi)^{3}}
(2\pi)\delta(p_z-q_z-k_z)
~T^{3}\sum_{l,n_1,n_2}~
\beta  \delta_{n_1\, ,\, n_2+l}
\frac{4~m_f^2}{[{\bf k}_L^2-m_k^2][{\bf p}_L^2-m_f^2][{\bf q}_L^2-m_f^2]}
\, ,
\label{excRF-d2-2}
\end{eqnarray}
and ${\bf k}_L=(i\omega_l^B,k_z)$, ${\bf p}_L=(i\omega_{n_1}^F,p_z)$, ${\bf p}_L=(i\omega_{n_2}^F,q_z)$. 
This expression realizes concretely the intuitive expectation that the nontrivial dynamics in a strong magnetic field is effectively $(1+1)$ dimensional\footnote{The transverse motion is not affected by the temperature, and is dictated by the lowest Landau level.  Increasing the magnetic field shrinks the 
individual  Landau orbits as $r \sim 1/\sqrt{q_{f}B}$. 
So, the orbital motion of quarks in the plane transverse to the magnetic field becomes more and more constrained 
as $B$ grows. In the limit of very large fields, the original helicoidal (tubular-like) paths become 
essentially straight lines parallel to the field direction. Of course, the longitudinal  motion is affected by the heat bath, so that the  tubular structures become ``blurred'' (noisy). However, if the 
magnetic field is larger than the temperature, this effect will be minor and the quark motion will be essentially one dimensional.}. Since the gluons do not couple directly to the magnetic field, their dispersion relation maintains its $(3+1)$-dimensional character ($\omega^2=k_1^2+k_2^2+k_3^2$), which effectively results in a ``massive'' gluon ($m_k^2=k_1^2+k_2^2$) in the reduced $(1+1)$-dimensional diagram. In the end the exchange contribution to the QCD pressure in the lowest-Landau level approximation for the fermion propagation is essentially an average over the effective gluon transverse mass $m_k^2=k_1^2+k_2^2$ of the exchange diagram in $(1+1)$-dimensions with the Gaussian weight $(q_fB/2\pi) \exp[-m_k^2/2q_fB]$.

Apart from the tensorial structure, this diagram corresponds to the two-dimensional version of the exchange diagram  of a Yukawa theory with both massive fermions and bosons, which was computed originally in Ref. \cite{Palhares:2008yq} (cf. also \cite{thesis}). In the present $(1+1)$-dimensional context, however, one expects renormalization to be trivial. It can be shown indeed that the usual quark self-energy counterterms vanish in dimensional regularization, while the high momenta in the gluon lines are fully tamed by the Gaussian weight $(q_fB/2\pi) \exp[-m_k^2/2q_fB]$.
Our concern here will be thus directed towards the IR domain, which proves to  be subtle as discussed for instance in 
Ref. \cite{Fukushima:2011nu}. 

Since the result in  Eq.~(\ref{excRF-LLL-4}) is (superficially) proportional to $m_f^2$ (see Eq.~\ref{excRF-d2-2}), 
it vanishes in the chiral limit $m_f\to 0$.  Simple power counting suggests that this result is not affected by the remaining integrations over the longitudinal momenta, which are are only logarithmically divergent when $m_f\to 0$. This is confirmed by a numerical analysis of the complete integral. This is calculated by  performing first the sums over the Matsubara frequencies, via standard contour integration. We obtain then,
for one massive flavor \cite{leticia-tese}

\vspace{0.3cm}

\begin{eqnarray}
\parbox{10mm}{
\fmfreuse{exchange}
}^{\quad\rm LLL}
 &=& 
\beta V~g^2N_c\left( \frac{N_c^2-1}{2} \right)
~
m_f^2
\left(
\frac{q_fB}{2\pi}\right)
\int\frac{d^2 {\bf k}_{\perp}}{(2\pi)^2}~{\rm e}^{
-\frac{{\bf k}_{\perp}^{2}}{2q_fB}}
~
\int \frac{d p_3 d q_3 d  k_3}{(2\pi)^{3}}
~(2\pi)\delta(k_3-p_3+q_3)
\nonumber\\
&&
\frac{1}{\omega\Ep\Eq}\Bigg\{
\frac{\omega~\Sigma_+}{E_-^2-\omega^2}
+
\frac{\omega~\Sigma_-}{E_+^2-\omega^2}
+2
\left[\frac{E_+}{E_+^2-\omega^2}-\frac{E_-}{E_-^2-\omega^2}\right]
~
n_B(\omega)~N_F(1)
-\nonumber \\
&&
-
\left[
\frac{2(\Eq+\omega)}{(E_--\omega)(E_++\omega)}
\right]
~N_F(1)
-2~
\frac{E_+}{E_+^2-\omega^2}
~n_B(\omega)
-
\frac{1}{E_++\omega}
\Bigg\}
\,,
 \label{exc-LLL-final}
\end{eqnarray}

\vspace{0.3cm}

\noindent
where $n_B$ and $n_F$ are the Bose-Einstein and Fermi-Dirac distributions, respectively, 
$\omega =\sqrt{k_3^2+{\bf k}_{\perp}^2}$, and 
\be
E_{\pm} &\equiv& \Ep\pm\Eq=\sqrt{{\bf p}_z^2+m_f^2}\pm\sqrt{{\bf q}_z^2+m_f^2} \, ,
\nonumber \\
N_F(1) &\equiv& n_F(\Ep+\mu_f ) +n_F(\Ep-\mu_f ) \, ,
\label{Nfs}
\\  
\Sigma_\pm &\equiv& n_F(\Ep+\mu_f )~n_F(\Eq\pm\mu_f )+n_F(\Ep-\mu_f )~n_F(\Eq\mp\mu_f ) \, .
\ee
\end{fmffile}
The integrals over ${\bf k}_{\perp}$ can be written in radial coordinates and put in a more convenient 
form by the change of variables $x\equiv |{\bf k}_{\perp}|/(2q_{f}B)$. The Gaussian weight becomes then
a representation of the Dirac delta function for very large magnetic fields (for $b\equiv 2q_{f}B$, 
$e^{-bx^{2}}/\sqrt{\pi/b}\to \delta(x)$), so that the UV limit is completely under control. By  
performing the integrals numerically, one finds that the overall $m_{f}^{2}$ factor controls indeed the IR sector: 
without this factor, the integrals would diverge in what seems to be a logarithmic fashion (as  can be also seen 
from semi-analytic stronger approximations in the limit of large fields). This confirms the result  anticipated in Eq. (\ref{excRF-d2-2}): {\it  the exchange contribution to the QCD pressure for very high magnetic fields vanishes in the chiral limit. }

It is clear, then, that the quark masses play an important role in the perturbative calculation of thermal QCD under an external magnetic background, competing with the field. In fact, although the effects from quark masses on the thermodynamics of QCD have been greatly overlooked for many years, it has been shown that they can bring significant corrections to the perturbative pressure at finite density \cite{Fraga:2004gz}, with consequences to the structure of compact stars, and at finite temperature \cite{Laine:2006cp}, affecting quark mass thresholds (see also \cite{Wang:1998tg}). In order to make direct comparisons to our results at large magnetic fields, we also present here an explicit analysis of the mass-dependence of the $O(g^2)$ thermal QCD pressure. 
The derivation follows again the same steps as the analogous computation for the massive Yukawa theory \cite{Palhares:2008yq} in dimension $4$, with the simplification that gluons are massless. For one-flavor QCD, after renormalization (in the $\overline{\rm MS}$ scheme), one obtains for the exchange contribution to the thermodynamical potential:
\be
\frac{\Omega_{QCD}^{F, (B=0)}}{N_c} &=&
-\frac{2}{\beta}~\int \frac{d^3{\bf p}}{(2\pi)^3} \left[ 
\ln\left( 1+e^{-\beta(\Ep-\mu_f)} \right)
+\ln\left( 1+e^{-\beta(\Ep+\mu_f)} \right)
\right]
-
\nonumber\\
\nonumber\\
&&
-~\frac{1}{2}~g^2\left( \frac{N_c^2-1}{2} \right)
\int \frac{d^3{\bf p} d^3{\bf q}}{(2\pi)^6}
\Bigg[
\frac{\mathcal{J}_{+}~\Sigma_+}{\Ep\Eq}
+
\frac{\mathcal{J}_{-}~\Sigma_-}{\Ep\Eq}
-4\frac{ n_B(\omega_{12})~N_F(1)}{\Ep\omega_{12}}
\Bigg]
+\nonumber \\ &&
+~\frac{1}{2}\frac{g^2}{(4\pi)^2}\left( \frac{N_c^2-1}{2} \right)
4m_f^2\left[ 4+6\log\left( \frac{\Lambda}{m_f} \right) \right]
\left[\int \frac{d^3{\bf p}}{(2\pi)^3}~
\frac{N_F(1)}{\Ep}\right]
\, ,
\ee
with $\Lambda$ being the $\overline{\rm MS}$ renormalization scale, $\omega_{12}^2\equiv ({\bf p}-{\bf q})^2$ and $\mathcal{J}_{\pm}\equiv -1-2m_f^2/(E_{\mp}^2-\omega_{12}^2)$.

\begin{figure}[!thb]
\includegraphics[width=0.5\textwidth]{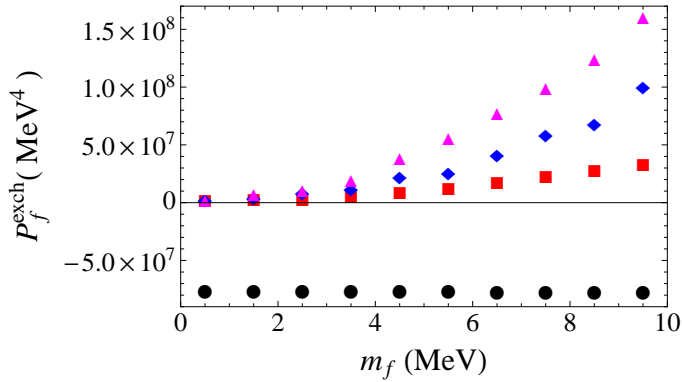}
\caption{Exchange contribution to the pressure, $P_{f}^{exch}$, as a function of the quark mass 
for $q_{f}=2/3$, $T=100$ MeV and $eB/m_{\pi}^{2}=0$ (lower), $100$, $200$ and $300$ (upper). The strong coupling is fixed at $\alpha_{s}\equiv g^2/4\pi = 0.3$ and the renormalization scale for the zero magnetic field curve is set to $\Lambda=800~$MeV.}
\label{plot-exchange-mass}
\end{figure}

In Figure \ref{plot-exchange-mass} we display the behavior of the contribution from the exchange diagram 
to the pressure as a function of  the quark mass for $q_{f}=2/3$, $T=100$ MeV and 
a few values of $eB$ in units of the pion mass, as well as for vanishing $eB$. The picture clearly shows that the exchange 
contribution vanishes smoothly in the chiral limit. As the quark mass is increased, the exchange pressure 
becomes nonzero but remains always between one and two orders of magnitude below the leading (free) 
contribution\footnote{A thorough comparison between the free pressure and the $O(\alpha_{s})$ contribution
as well as thermal effects will be presented in a longer publication.}.
This fact might indicate an improved convergence of the perturbative series in the quark sector\footnote{Notice that the purely gluonic sector remains unaffected, so that it is obviously still plagued by infrared divergences as its $B=0$ analog. Nevertheless, for extremely large fields, the contribution of pure-gluon diagrams to the pressure (which are $B$-independent) can be neglected as a subleading correction and the QCD pressure will be dominated by the quark sector.} in this limit of high temperatures and 
extremely large magnetic fields. This is further supported by the absence of any explicit dependence of the pressure on the 
renormalization scale. Higher-loop computations or a direct comparison with lattice QCD data for 
the pressure could confirm this in the near future.

The comparison with the behavior of the pressure at zero magnetic fields shows that the exchange contribution to the pressure changes its sign when the QCD medium is exposed to a very intense magnetic background,  $\sqrt{eB}\gg T$. This qualitative change is directly related to the effective dimensional reduction at large magnetic field and the consequent modification of the Dirac traces (the vanishing of the exchange pressure in the chiral limit can be understood as resulting from a conflict arising, in one dimension, from the fact that the exchange diagram couples right and left movers, while the vertices conserve helicity). The nonzero effective mass acquired by the gluon in the dimensionally-reduced diagram also contributes to prevent cancellations that usually happen at zero magnetic fields.

We would like to conclude with an intriguing observation. 
It is well-known that the gyromagnetic factor of Dirac fermions, $g_{m}$, is not exactly $2$, but that it receives radiative corrections from QED (also from QCD,  but these are subleading). The deviation, although small ($g-2\sim 10^{-3}$) may produce sizable corrections to the pressure. Indeed, it affects the energy of the lowest Landau level, effectively turning the mass $m$  into $m_{eff}^2=m^2 +(g-2) eB$. For $eB\sim m_\pi^2$, the correction is in the MeV range; it may thus compete with $m$ and cannot be ignored in a quantitative study. \footnote{Of course, the correction in $(g-2)$ will be affected by $B$ and $T$, so that it is not clear what will happen with its sign and magnitude at large fields. At zero temperature, there seems to be a non-monotonic behavior \cite{Baier:1975uj}.} 

\section*{Acknowledgments} 
We thank M. Chernodub, G. Dunne, G. Endr\H odi, A. Peshier and I. Shovkovy for useful discussions. 
L.F.P. is grateful to the hospitality of IPhT, during a long-term visit, where part of this work has been done. 
E.S.F. thanks the ECT$^{*}$ for the hospitality during the workshop {\it QCD in strong magnetic fields} where helpful discussions took place.
This work was partially supported by CAPES, CNPq, FAPERJ and FUJB/UFRJ. LFP acknowledges the support of the Alexander von Humboldt Foundation. JPB acknowledges the support of  the European Research Council under the Advanced Investigator Grant ERC-AD-267258.



\end{document}